\documentclass[12pt]{article}
\usepackage{graphics, color}
\usepackage{graphicx}
\usepackage{amssymb}
\pdfoutput=1

\newcommand{\sect}[1]{\section{#1}\setcounter{equation}{0}}

\def\gsim{\, \rlap{$>$}{\lower 1.1ex\hbox{$\sim$}}\,}
\def\lsim{\, \rlap{$<$}{\lower 1.1ex\hbox{$\sim$}}\,}
\newcommand{\vk}{\vec{k}}

\begin{document}

%Title page

\begin{titlepage}
\rightline{ NSF-KITP-10-010}
\bigskip
\bigskip\bigskip\bigskip
\centerline{\Large Semi-Holographic Fermi Liquids}
\bigskip\bigskip\bigskip
\bigskip\bigskip\bigskip

 \centerline{{\bf Thomas Faulkner}\footnote{\tt faulkner@kitp.ucsb.edu}
and {\bf Joseph Polchinski}\footnote{\tt joep@kitp.ucsb.edu}}
\medskip
\centerline{\em Kavli Institute for Theoretical Physics}
\centerline{\em University of California}
\centerline{\em Santa Barbara, CA 93106-4030}\bigskip
\bigskip
\bigskip\bigskip

%ABSTRACT

\begin{abstract}
We show that the universal physics of recent holographic non-Fermi liquid models is captured by a semi-holographic description, in which a dynamical boundary field is coupled to a strongly coupled conformal sector having a gravity dual.  This allows various generalizations, such as a dynamical exponent and lattice and impurity effects.  We examine possible relevant deformations, including multi-trace terms and spin-orbit effects.  We discuss the matching onto the UV theory of the earlier work, and an alternate description in which the boundary field is integrated out.
\end{abstract}
\end{titlepage}
\baselineskip = 17pt
\setcounter{footnote}{0}

\sect{Introduction}                   

The existence of non-Fermi liquids, conductors whose gapless charged excitations are not described by the Landau-Fermi liquid effective theory, is a fascinating puzzle in condensed matter physics.\footnote{For a recent overview of this subject see Ref.~\cite{senthil}.}  Even a clear framework is lacking.  One approach, the marginal Fermi liquid~\cite{Varma:1989zz}, is a long-standing phenomenology without clear field-theoretic underpinnings.  Another, the recent attempt to formulate scaling laws~\cite{senthil}, places this subject at the level of development of critical phenomena before the advent of the renormalization group.  A third class of ideas, based on emergent gauge theories, presents the difficulty of understanding the resulting strongly coupled dynamics~\cite{leeFL}.

In this situation, gauge/gravity duality may play a valuable role.  Having a large class of solvable quantum field theories provides a theoretical laboratory for understanding phenomena that emerge at strong coupling.  This has led to a recent surge of interest in studying duals that capture various features of condensed matter systems; for reviews see Ref.~\cite{reviews}.

The known non-Fermi liquids retain a Fermi surface, a surface in momentum space where the electron propagator has IR-singular behavior.  In this paper we develop further the duals introduced in Refs.~\cite{Lee:2008xf,Liu:2009dm,Cubrovic:2009ye,Faulkner:2009wj}, which exhibit such a Fermi surface.\footnote{Ref.~\cite{KulaxiziParnachev} identifies duals in which a current-current propagator exhibits Fermi-surface-like behavior, although backreaction effects may prevent this from extending fully into the IR~\cite{HPST}.  In these duals the sea fermions would be charged under the large-$N$ gauge group, whereas in the systems that we consider they are gauge singlets.  Ref.~\cite{Rey:2008zz} identifies some properties of a charged AdS black hole with those of a Fermi gas, but there are no indications of a Fermi surface in the low frequency correlators.}  In particular, we show that the IR behavior found in Refs.~\cite{Liu:2009dm,Faulkner:2009wj} is equivalent to that in a system of dynamical singlet fermions coupled through a fermion bilinear to a strongly coupled conformal sector with  $AdS_2$ dual.  We argue that this semi-holographic description has significant advantages: it retains only the universal low energy properties, which are most likely to be relevant to the realistic systems; it allows more flexible model-building; and, it makes it easy to incorporate such features as a spatial lattice and impurities.   As an example of the flexibility, we consider the replacement of the holographic $AdS_2$ sector by systems with other $AdS$ and Lifshitz scalings.

We present this construction in Sec.~2.  In Sec.~3 we look at the relevant perturbations of this system, with regard to stability and to the phase diagram.  There are a large number of relevant perturbations, multi-trace operators constructed out of the basic fields and currents.  These do not appear to destabilize the construction in general.  Particular operators, the squares of densities, seem interesting for the phase diagram.  We also discuss spin-orbit effects.  In Sec.~4 we discuss the matching of the IR theory onto the UV theory of Refs.~\cite{Liu:2009dm,Faulkner:2009wj}, and we note an interesting renormalization group interpretation of the Fermi surface.

\sect{Semi-holographic Fermi liquids}

Refs.~\cite{Lee:2008xf,Liu:2009dm,Cubrovic:2009ye,Faulkner:2009wj} consider 2+1 dimensional conformal theories with a density of conserved charge, whose dual description is a charged $AdS_4$ black hole.  At zero temperature the near-horizon geometry is $AdS_2 \times {\bf R}^2$.  In addition there is a charged fermion in the bulk, dual to a charged fermionic operator.  In Refs.~\cite{Liu:2009dm,Faulkner:2009wj}, the Fermi surface of the field theory arises from a sea of these bulk fermions, radially localized on the domain wall separating the UV $AdS_4$ and IR $AdS_2 \times {\bf R}^2$ geometries.

\begin{figure}
\begin{center}
\includegraphics[scale=1]{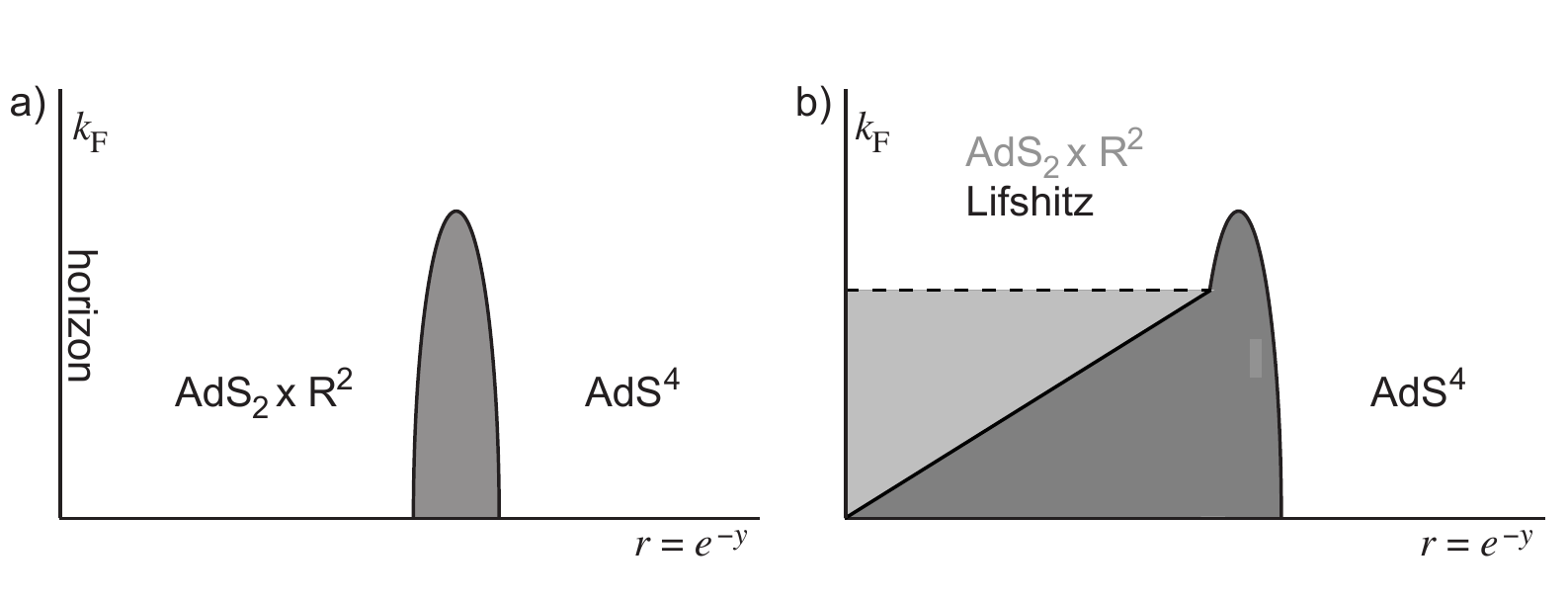} 
\caption{Fermi momentum as a function of radius, shown in the WKB approximation.
a) The minimal ingredients to give rise to non-Fermi liquid behavior, a domain wall Fermi sea plus $AdS_2$ horizon.  b) Over most of the parameter space of Ref.~\cite{Faulkner:2009wj} there is also a Fermi liquid in the IR bulk (entire shaded region).  Backreaction converts the geometry to Lifshitz form~\cite{HPST}, whose Fermi sea is shown in dark shading.}
\label{fig:domain}
\end{center}
\end{figure}

This is interesting from the holographic point of view.  The essential low energy degrees of freedom of the system consist of the excitations of the bulk Fermi surface, localized at the domain wall, plus the holographic excitations at the $AdS_2 \times {\bf R}^2$ horizon (Fig.~\ref{fig:domain}a).  There is also a model-dependent Fermi liquid in the bulk of $AdS_2 \times {\bf R}^2$ (Fig.~\ref{fig:domain}b), but this is not connected with the non-Fermi liquid behavior of Refs.~\cite{Liu:2009dm,Faulkner:2009wj}, so we discuss it later.
The system shown in Fig.~\ref{fig:domain}a is in the same universality class as a rather different quantum field theory, consisting of free fermions coupled to whatever CFT is dual to the $AdS_2$ bulk theory.
The action is\footnote{A similar construction was noted in \cite{Faulkner:2009wj}
but not elaborated upon.}
\begin{equation}
S =  S_{\rm strong} + \sum_s \int dt\int \frac{d^2 k}{(2\pi)^2} \, \left\{ \chi_{\vec k,s}^\dagger (i \partial_t -  \varepsilon_{\vec k} + \mu) \chi^{\vphantom\dagger}_{\vec k,s}
+ g^{\vphantom *}_{\vec k} \chi_{\vec k,s}^\dagger \Psi^{\vphantom\dagger}_{\vec k,s} + g_{\vec k}^*  \Psi^{\dagger}_{\vec k,s} \chi_{\vec k,s}^{\vphantom\dagger} \right\}\,.
\label{action}
\end{equation}
Here $\chi_{\vec k,s}^{\vphantom\dagger}$ is a charged singlet operator, free aside from the explicit coupling in $S$,
$\varepsilon_{\vec k}$ is the single-particle energy of the singlet fermions, $\mu$ is the chemical potential, and $g_{\vec k}$ is an arbitrary coupling function.  Also, $\Psi^{\vphantom\dagger}_{\vec k,s}$ is a charged fermionic operator from the strongly coupled sector, of dimension $\Delta_{\vec k}$, with $s$ the spin component.
Only the leading behaviors near the Fermi surface are relevant,
\begin{equation}
\varepsilon_{\vec k} - \mu \approx v_{\vec k_{\star}} k_\parallel\ , \quad g_{\vec k} \approx g_{\vec k_{\star}}\ , \quad \Delta_{\vec k} \approx \Delta_{\vec k_{\star}}\, ,
\label{linear}
\end{equation}
where $\vec k_{\star}$ is the point on the Fermi surface nearest to $\vec k$, and $k_\parallel
= |\vec k - \vec k_{\star}|$.

We refer to this construction as semi-holographic:
the correlators of the strongly coupled theory are to be calculated via AdS/CFT duality, and then the
singlet field integrated explicitly.  The singlet couples to local operators in the strongly coupled sector, so is naturally regarded as living on the boundary.\footnote{Dynamical boundary fields were introduced in Ref.~\cite{Witten:2003ya}.  Some further developments of this idea are in Refs.~\cite{dyn}. }
This construction gives a general framework for analyzing the low energy physics of the systems of Refs.~\cite{Lee:2008xf,Liu:2009dm,Cubrovic:2009ye,Faulkner:2009wj}, while dropping nonuniversal physics.   Unlike the phenomenological constructions of bulk duals, here there is an immediate top-down interpretation: given any known dual with a $U(1)$ symmetry and charged fermionic operators, we can couple an additional singlet fermion in this way.  Further, we are free to consider other scaling behaviors in the strongly coupled sector, such as $AdS_4$ and Lifshitz spacetimes.  

There is no advantage to generating the Fermi sea in the bulk of a larger holographic theory. (In particular, any desire to obtain the $\chi$ fields from an explicit UV brane should be resisted.)  Since these fermions are gauge singlets, the geometry does no work for us in summing their quantum effects; if these effects are important we must deal with them ourselves.  Further, obtaining both sectors holographically is complicated and constraining.  These complications may be essential in determining the low energy physics of a given theory (if a holographic description holds above the scale of the Fermi momentum), but to classify all possible low energy physics within a given universality class the semi-holographic description is more efficient. 

As one example of the flexibility of the semi-holographic construction, we can immediately introduce the effects of a spacetime lattice, simply by restricting the momentum $\vec k$ of $\chi$ to a unit cell of the reciprocal lattice, with $\epsilon_{\vec k} = \epsilon_{\vec k+\vec K}$ for any reciprocal lattice vector $\vec K$.\footnote{For a cubic lattice of side $a$, this restricts the momentum components to $-\pi/a < k_i < \pi/a$ with periodic boundary conditions.}  The coupling to the strongly coupled sector must be generalized to
\begin{equation}
\label{umk}
\sum_{\vec K} \chi_{\vec k,s}^\dagger \Psi^{\vphantom\dagger}_{\vec k+\vec K,s} + {\rm h.c.}\, .
\end{equation}
This allows transfer of momentum $\vec K$ to the lattice, as is essential to the DC resistivity~\cite{AshMer}.  Similarly, we can introduce impurities, by adding to the action $\chi^\dagger\chi$ and/or $\Psi^\dagger\chi + \chi^\dagger\Psi$ terms that violate momentum conservation.\footnote{For other approaches to lattices and impurities see Refs.~\cite{lattice,dirt} respectively.}

To leading order in $1/N$, the action for the bulk fermion  $\psi_{\vec k,s}^{\vphantom\dagger}$ dual to 
the operator $\Psi_{\vec k,s}^{\vphantom\dagger}$ is quadratic, and so the coupled action for $\psi$ and $\chi$ is quadratic.  Thus we can immediately calculate the correlators.  In fact, we can do this directly in the quantum theory, though the bulk description would be needed to obtain transport and thermal properties.     We use $G$ and $\cal G$ respectively for the $\chi$ and $\Psi$ correlators.
The decoupled Feynman correlators are
\begin{equation}
G_0 (\vec k, \omega) = \frac{1}{\omega - v_{\vec k_*} k_\parallel}\, ,\qquad {\cal G}_0(\vec k,\omega)  = c_{\vec k} \omega^{2\Delta_{\vec k}-1}\, ,
\end{equation}
where the subscripts indicate that $g_{\vec k}$ has been set to zero.  The $\omega$ dependence of ${\cal G}_0$ is determined by scale invariance; this becomes $\omega\ln\omega$ at $\Delta_{\vec k} = 1$.  The coefficient $c_{\vec k}$ is obtained by a holographic calculation in the IR theory~\cite{Faulkner:2009wj,FILMV} for the case of $AdS_2 \times {\bf R}^2$ with an electric field in $AdS_2$.   The magnitude of $c_{\vec k}$ 
depends on convention, but there is phase which is physical.\footnote{We are neglecting possible spin-orbit coupling terms in the action (\ref{action}), so as not to distract from the main point of the following discussion.   In Sec.~3.3 we introduce these terms, which are necessary to match to the original $AdS_4$ model. Because of these spin-orbit effects $\epsilon_{\vk}$ and $c_{\vec k}$ may be matrices in the spin basis, which we suppress for now.}

\begin{figure}
\begin{center}
\includegraphics[scale=1.1]{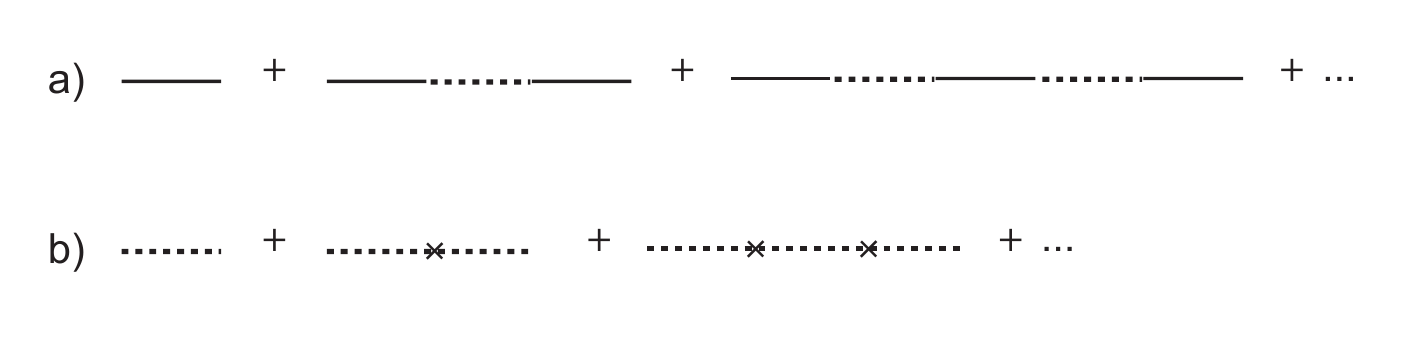} 
\caption{a) The geometric sum leading to the fermion correlator~(\ref{gcorr}).  The solid line represents $G_0$ and the dashed line represents ${\cal G}_0$.  b) The geometric sum~({\ref{gof}}), where an $\times$ represents the double-trace perturbation.}
\label{fig:geo}  
\end{center}
\end{figure}

Expanding in powers of $g_{\vec k}$, and using the factorization property of large-$N$ correlators, the $\chi$ propagator is (Fig.~\ref{fig:geo})
\begin{eqnarray}
G_g (\vec k, \omega) &=& \sum_{n=0}^\infty  |g_{\vec k}|^{2n} G_0 (\vec k, \omega)^{n+1} {\cal G}_0(\vec k,\omega)^n 
\nonumber\\
&=& \frac{1}{ G_0 (\vec k, \omega)^{-1} -  |g_{\vec k}|^{2}  {\cal G}_0(\vec k,\omega)}
\nonumber\\
&=& \frac{1}{\omega - v_{\vec k_*}  k_\parallel - c_{\vec k} |g_{\vec k}|^{2} \omega^{2\Delta_{\vec k}-1} }
\, , \label{gcorr}
\end{eqnarray}
exhibiting the strange metallic behavior discussed in Refs.~\cite{Liu:2009dm,Faulkner:2009wj}.  The calculation uses only the factorization property and so would be equally true in weakly coupled large-$N$ theories.

Now we can see what happens if the $AdS_2 \times {\bf R}^2$ strongly coupled theory is replaced by an $AdS_4$ theory, with $\Psi$ an operator of dimension $\Delta$ in the $2+1$ dimensional CFT.  The correlator 
\begin{equation}
\langle 0 | {\rm T}\, \Psi(\vec x,t) \Psi^\dagger(0,0) | 0 \rangle = (x^2 - t^2)^{-\Delta}
\end{equation}
implies
\begin{equation}
{\cal G}_0(\vec k,\omega) = A(\Delta) (k^2 - \omega^2 )^{\Delta - 3/2}\, ,
\end{equation}
with $A(\Delta) = 4\pi  \Gamma(2-2\Delta) \sin{\pi\Delta}$.  Since the Fermi momentum $\vec k$ is a UV scale we are interested in $k \gg \omega$ and so we expand,
\begin{equation}
{\cal G}_0(\vec k,\omega)  = A(\Delta) \left\{ k^{2\Delta - 3} - (\Delta - 3/2) \omega^2 k^{2\Delta - 5} + \ldots \right\}\, .
\end{equation}
Using this in the correlator~(\ref{gcorr}), the leading $\omega$-independent term should be absorbed into the definiton of $\varepsilon_{\vec k}$ from the UV theory.  The leading correction to the fermion self-energy is $\omega^2$ as in Fermi liquid theory, but here it is real because the kinematics forbids the quasiparticle decay. This example should capture
the low energy dynamics of the models considered in \cite{Faulkner:2009am} where a 
fermion lives in a zero temperature holographic superconducting background
and the IR part of the geometry was an emergent $AdS_4$ solution.

Similarly, we can extend to a Lifshitz theory with dynamical exponent $z$.  For an operator of energy dimension $\Delta$,
\begin{equation}
{\cal G}_{0}(\vec k,\omega)  = A(\Delta,z) k^{2\Delta/z - 2 - z} + B(\Delta,z) \omega^2 k^{2\Delta/z - 2 - 3z} + \ldots  \, .\label{lifex}
\end{equation}
This asymptotic expansion in $\omega^2$ is seen, for example, in the holographic calculation of the correlator~\cite{Kachru:2008yh}, obtained by solving the differential equation
\begin{equation}
-\partial_y^2 \phi'  + (\Delta - \beta)^2  \phi' + k^2 e^{2y/z} \phi' - \omega^2 e^{2y} \phi' = 0\, .
\label{wave}
\end{equation}
Here we have for simplicity taken a scalar correlator to illustrate the scaling, in the metric 
\begin{equation}
ds^2 = L^2(dy^2 + e^{-2y/z} d\vec{x}^2 - e^{-2y} dt^2)\, ,\label{metric}
\end{equation}
 and we have defined $\phi' = e^{\beta y}\phi$, $\beta = (z+2)/2z$.   
At $\omega=0$, the $k^2$ term is dominant at the horizon, giving the normalizable solution $\phi' \sim \exp\left( - k z e^{y/z} \right)$.  Perturbation theory in $\omega^2$ is nonsingular order by order because this solution falls off rapidly at large $y$, 
so again the leading correction scales as in Fermi liquid theory, but with a real  coefficient.    

The fermionic correlator will have similar scaling properties.  Thus, the non-Fermi liquid behavior found in Refs.~\cite{Liu:2009dm,Faulkner:2009wj} is only present for the $AdS_2 \times {\bf R}^2$ IR theory, in which the momentum does not scale.  The fermionic correlator would have interesting properties when $k$ and $\omega$ are both small, but this does not affect the Fermi surface behavior.  Also, when there is bulk Fermi sea as in Fig.~\ref{fig:domain}b, the dimension $\Delta$ becomes complex, but again this is seen only when $k$ and $\omega$ become small together.

Let us discuss a few more aspects of the Lifshitz case.  For finite values of $\omega$, the $\omega^2$ term dominates at the horizon and the boundary conditions are changed to $\phi' \sim \exp\left( \pm i \omega e^{y} \right)$.  The quasiparticle width is a tunneling effect and can be estimated
using the WKB approximation, $\Sigma \sim \exp(-\Gamma)$ where
\begin{equation}
\Gamma \sim 2 \int_{-\log(\mu)}^{y_{\rm turn}} dy\, \sqrt{(\Delta - \beta)^2 +  k^2 e^{2 y/z} - \omega^2 e^{2 y} } 
\, ,
\label{Gamma}
\end{equation}
and $\mu$ is a UV cutoff.  At small frequency the barrier is large, and we can approximate $y_{\rm turn} = \frac{z}{2(z-1)} \log(k^2/\omega^2)$. We can neglect the $\Delta -\beta$ terms
and send the UV cutoff to infinity to find,
\begin{equation}
\Gamma \sim 2 \int_{-\infty}^{y_{\rm turn}} dy \sqrt{k^2 e^{2 y/z} - \omega^2 e^{2 y} } 
 =  \left( \frac{k^z}{\omega} \right)^{\frac{1}{z-1} }  \frac{\sqrt{\pi}\, \Gamma\left(\frac{1}{2(z-1)} \right)}{
  \, \Gamma\left( \frac{z}{2(z-1)} \right)}
\, .
\label{lifscale}
\end{equation}
The imaginary part goes to zero exponentially at low frequency. 

It is interesting to examine the crossover from Lifshitz to $AdS_2$ behavior as $z$ becomes large ($AdS_2$ corresponding to $z = \infty$). To do this
we expand around the answer for $z=\infty$, which can also be read
off from the WKB approximation:
\begin{equation}
\Gamma \sim 2 \int^{y_{\rm turn}}_{-\log(\mu)}
dy \sqrt{ \nu_{\vec k}^2 - \omega^2 e^{2 y} + 2 k^2 y  /z} 
\end{equation}
where $\nu_{\vec k}^2 = k^2 + (\Delta-\beta)^2$. For large
$z$ the turning point is $y_{\rm turn} = \log(\nu_{\vec k}/\omega)$.
We can expand the integrand in $z$ as long
as the $1/z$ term is subdominant over the domain
of the integral. This is the case for $\omega > \nu_{\vec k} \exp(- 2 z \nu_{\vec k}^2 / k^2)$
such that now the quasi particle width behaves as,
\begin{equation}
\Sigma \sim (\omega/\mu)^{2 \nu_{\vec k}} \left( 1 - \frac{k^2}{2 z \nu_{\vec k}^2} \log(\omega/\mu) + \ldots \right) 
\end{equation}
We have recovered the $AdS_2$ scaling with log corrections and
scaling dimension $\Delta_{\vec k} = \nu_{\vec k} + 1/2$.
The $\log$'s will be small for $\nu_{\vec k} \exp(- 2 z \nu_{\vec k}^2 / k^2) < \omega
\ll \mu$.
For energy scales $\omega < \nu_{\vec k} \exp(- 2 z \nu_{\vec k}^2 / k^2)$ the answer will then
cross over to (\ref{lifscale}). 
Thus for modestly large values of $z$ the strange metal behavior extends down to very low energies with small logarithmic corrections.

The Lifshitz case studied above should apply to the low energy
dynamics of fermions introduced into various known gravity backgrounds with
Lifshitz as the IR fixed point. Examples are the charged dilaton black
holes considered in \cite{Goldstein:2009cv} and the extremal
limit of some holographic superconductors \cite{Gubser:2009cg},
though some details will be slightly
different. For example the quasiparticle
width will now depend on the charge of the fermion due to the radial electric field in these
backgrounds. We expect the $z\rightarrow \infty$ limit of  the models in \cite{Goldstein:2009cv} will
recover many of the details of the $AdS_2 \times {\bf R}^2$ theory including
the scaling exponent given in Eq.~(\ref{exponent}), hence they deserve more detailed study. 

When the dimensions $\Delta_{\vec k}$ in the $AdS_2 \times {\bf R}^2$ case lie partly on the conformal branch where $2\Delta_{\vec k} - 1$ is imaginary, a Fermi sea also develops in the $AdS_2$ region.  The backreaction from this sea changes the geometry to the Lifshitz form, with a dynamical exponent that is large in the planar limit of the strongly coupled theory~\cite{HPST}.  Over most of the parameter space of Ref.~\cite{Faulkner:2009wj} this low energy Fermi sea accompanies the domain wall Fermi sea, though there is a small range of parameters where the latter is absent (this will be discussed in Section 4 in more detail).  From the semi-holographic point of view here, the bulk and boundary Fermi seas are independent features.

The marginal Fermi liquid phenomenology~\cite{Varma:1989zz} also involves a Fermi sea coupled to an $AdS_2 \times {\bf R}^2$-like sector, the latter having correlators that are singular at $\omega = 0$ for all $\vec k$.  In the present work, however, the two sectors are coupled through a fermion mixing interaction, whereas in the marginal Fermi liquid they are coupled through bosonic operators.

In the strongly coupled CFT, the operator $\Psi$ is a composite of the fundamental fields.  Thus the semi-holographic Fermi liquid has a natural interpretation describing charge carriers that have some amplitude to be pointlike and some amplitude to fragment.  Such fragmentation plays a major role in many attempts to understand exotic phases~\cite{Anderson:1997vm}.  It is interesting that in the present construction this fragmentation is only possible with $AdS_2$ scaling.  The electrons near the Fermi surface have large momentum, and so, except in the case that momentum does not scale, their fragments are far off shell and cannot separate substantially.

If the electron does fragment, it is natural to ask what it fragments into.  In the supergravity limit there is no answer to this question, because the field theory is very strongly coupled (large 't Hooft coupling).\footnote{The term `quasiparticle' is used in condensed matter for any particle-like excitation in a low energy effective theory, like the charge carriers in Fermi liquid theory.  The term `unparticle' (or, better, `unparticle stuff') is used in some circles~\cite{Georgi:2007ek} for the excitations in strongly coupled scale invariant theories.  Thus one could say that the electron fragments into `unquasiparticles,'  or perhaps `quasiunparticles.'  }  This is one of the difficulties in applying gauge/gravity duality to QCD: the partonic structure is not seen.  Fortunately there is a wealth of phenomena in heavy ion physics where the partons are not relevant and the 't Hooft coupling is of order one.   In the condensed matter case, in the 2+1 dimensional systems where fragmentation is most often considered, the gauge coupling again is of order one.  The details of the partons may therefore not be so important --- indeed, there may be several equally valid dual descriptions.  Thus the situation may not be so different from that in heavy ion physics.

Having a fundamental fermion $\chi$ coupled to the composite $\Psi$ would appear to be a new ingredient outside the usual fragmentation picture.  In Sec.~3.1 we will show that in the interesting range of parameters $\chi$ can be integrated out.

\sect{Relevant operators}

Relevant operators are important for two reasons: they can make the low energy physics unstable to generic quantum corrections, and they affect the phase diagram as parameters are varied.  The phase diagram of the cuprates appears to be controlled by a single relevant operator, tied to the doping.\footnote{The doping appears directly in the action~(\ref{action}) through the chemical potential $\mu$, but integrating out the high energy degrees of freedom will make all other parameters implicit functions of $\mu$.  Writing the action~(\ref{action}) explicitly in terms of $\varepsilon_{\vec k} -  \mu$ is therefore somewhat deceptive.  Rather, the low energy kinetic term depends on the locus of $\vec k_\star$ (the Fermi surface) and on the normal derivative of the energy, $v_{\vec k_*}$, both of which are functions of $\mu$. We will
see this explicitly when mapping to the $AdS_4$ theory in section 4.}   At large doping $x$, the zero temperature behavior is that of a Fermi liquid, and non-Fermi liquid behavior sets in at a temperature $T_{\rm FL}$.  As the doping is reduced to a critical value $x_{\rm c}$, $T_{\rm FL}$ comes down close to zero (when the superconducting phase is suppressed).  Below $x_{\rm c}$ there is a mysterious phase with possible charge and spin inhomogeneities.  

As we will see, the theory we are discussing necessarily has an abundance of relevant operators.  However, the low energy physics is stable, within a range of parameters.  We also identify operators which may play the needed role in the phase diagram.

\subsection{Fermionic bilinears}

Candidate operators are the various bilinears, $\chi_{\vec k,s}^\dagger \chi^{\vphantom\dagger}_{\vec k,s}$,  $\Psi^{\dagger}_{\vec k,s} \chi_{\vec k,s}^{\vphantom\dagger}$, $\chi_{\vec k,s}^\dagger \Psi^{\vphantom\dagger}_{\vec k,s}$, and  $\Psi^{\dagger}_{\vec k,s} \Psi_{\vec k,s}^{\vphantom\dagger}$.  

From the action~(\ref{action}) it follows that $\chi_{\vec k,s}^\dagger \chi^{\vphantom\dagger}_{\vec k,s}$ has energy dimension zero and so is relevant.  However, this term is already present in the action: perturbing its coefficient just moves the Fermi surface, but does not destroy it.  Similarly the operators $\Psi^{\dagger}_{\vec k,s} \chi_{\vec k,s}^{\vphantom\dagger}$ and $\chi_{\vec k,s}^\dagger \Psi^{\vphantom\dagger}_{\vec k,s}$ are already included with generic coefficients.  Their dimension is $\Delta_{\vec k}$, and so they are relevant for $\Delta_{\vec k} <  1$.  This is just the regime where the correlator~(\ref{gcorr}) exhibits strange behavior (the marginal case is also of interest).

For genericity we should also add
\begin{equation}
S \to S - \sum_s \int dt\int \frac{d^2 k}{(2\pi)^2} \, f_{\vec k} \Psi_{\vec k,s}^\dagger \Psi^{\vphantom\dagger}_{\vec k,s}\,.
\end{equation}
This is relevant when $\Delta_{\vec k} < \frac{1}{2}$.  Such double-trace perturbations have been considered in Refs.~\cite{Aharony:2001pa,Witten:2001ua,Berkooz:2002ug,Gubser:2002zh,Gubser:2002vv}.
To see their effect, consider first the decoupled $\Psi$ theory at $g_{\vec k} = 0$.  As in the calculation of the propagator~(\ref{gcorr}), we can sum the perturbations at planar order to get
\begin{equation}
{\cal G}_{0,f}(\vec k,\omega)  = \frac{{\cal G}_{0}(\vec k,\omega)  }{1 - f_{\vk} {\cal G}_{0}(\vec k,\omega) }\,,
\label{gof}
\end{equation}
where again ${\cal G}_0(\vec k,\omega)  = c_{\vec k} \omega^{2\Delta_{\vec k}-1}$.  In the irrelevant range $\Delta_{\vec k} > \frac{1}{2}$, the perturbed correlator ${\cal G}_{0,f}$ approaches ${\cal G}_{0}$ at low energy.  However, for $\Delta_{\vec k} < \frac{1}{2}$ the low energy limit is 
\begin{equation}
{\cal G}_{0,f}(\vec k,\omega)  \approx  \frac{1}{f^{\vphantom l}_{\vk}} +  \frac{1}{ f_{\vk}^2 {\cal G}_{0}(\vec k,\omega) } + \ldots\,.
\end{equation}
The first term is a contact interaction and can be absorbed into the parameters in the action.  The second, which is the leading nonlocal term, scales as $\omega^{1 - 2\Delta_{\vec k}}$.  The relevant perturbations thus leave a critical theory but shift the dimension to $\Delta'_{\vec k} = 1 - \Delta_{\vec k}$~\cite{Witten:2001ua,Klebanov:1999tb}.  The fine-tuned theory with which we began is known as the alternate quantization, and it flows to the standard quantization.  The low energy physics is the same as would have been obtained using $\Delta'_{\vec k}$, the standard quantization, from the start.

In summary, the effect of generic fermionic bilinears is simply to restrict the dimensions to $\Delta_{\vec k} \geq \frac{1}{2}$, but otherwise leave the low energy physics as before.  These bilinears do not seem suitable for controlling the behavior of the phase diagram.  In particular, the different directions on the Fermi surface do not talk to each other, at least in the planar limit, and their coefficients are independent.  Thus it is difficult to see how they can account for a single relevant operator that drives the entire Fermi surface critical at once, as indicated by the existence of a sharp $x_{\rm c}$.

We present one more manipulation along these lines, which gives an interesting interpretation to the non-Fermi surface.  We can integrate out the singlet fermion $\chi$ and express the result as a boundary condition on the bulk theory.  This would usually be nonlocal, because of the time derivative term in the action.  However, when $\Delta_{\vec k} <  1$ one sees from the propagator~(\ref{gcorr}) that the time derivative is is less relevant than the interaction with the bulk.  We can therefore ignore it at low energy, so that $\chi$ is effectively nondynamical.  Integrating $\chi$ out then gives the action
\begin{equation}
\label{fm}
S' =  S_{\rm strong} - \sum_s \int dt\int \frac{d^2 k}{(2\pi)^2} \, \frac{| g_{\vec k}^2|}{\mu - \varepsilon_{\vec k}}\Psi_{\vec k,s}^\dagger \Psi^{\vphantom\dagger}_{\vec k,s} \,.
\label{intout}
\end{equation}
Assuming that we start with the standard quantization, the pole implies that on the Fermi surface we have transformed to the alternate quantization.  To see this, we evaluate the $\Psi \Psi^\dagger$ correlator by using the result (\ref{gof}) with $f_{\vk}$ given by Eq.~(\ref{intout}):
\begin{equation}
{\cal G}_{0,f}(\vec k,\omega)  = \frac{{\cal G}_{0}(\vec k,\omega)  }{1 - | g_{\vec k}^2| {\cal G}_{0}(\vec k,\omega) /(\mu -  \varepsilon_{\vec k})}\,.
\label{intout2}
\end{equation}
The second term in the denominator scales as $\omega^{2\Delta_{\vk} - 1} / k_\parallel$.  Thus, for $\omega > O(k_\parallel)^{\frac{1}{2\Delta_{\vk} - 1}}$ this term dominates and the propagator is approximately given by a contact term plus a term that has the alternate scaling.  Below this frequency the correlator crosses over to the standard scaling.  As we approach the Fermi surface the crossover frequency goes to zero and the alternate quantization applies down to the IR (though with an overall normalization on the correlator that goes to zero). See
also Fig. \ref{fig:1} and the discussion at the end of Section 4.1. 

\subsection{Density multilinears}

In this paper we are considering only universal properties of theories whose IR physics includes the system~(\ref{action}).  In addition to the fermionic operators $\Psi$, this will include the energy momentum tensor and the $U(1)$  density $j^t$ and current  $\vec \jmath$, under which $\Psi$ is charged.
In the $AdS_2$ theory, the density $j^t$ has dimension zero: the dual CFT is 0+1 dimensional, and so the current has the same scaling as the charge.  The spatial derivatives also do not scale.  Therefore {\it any function} of $j^t$ and $\vec k$ is relevant~\cite{HPST}.  This is an embarrassment of riches, and may lead one to doubt that an $AdS_2$ fixed point can ever be realized.  However, in the planar limit the RG flows induced by multi-trace operators (functions of gauge-singlet operators) are rather tame~\cite{Witten:2001ua,Berkooz:2002ug,Gubser:2002zh,Gubser:2002vv}.\footnote{These perturbations are nonlocal in the compact directions of the bulk dual~\cite{Aharony:2001pa}, but this does not lead to any pathologies.} The theory flows to a new fixed point with the same conformal symmetry, and only the correlator of the perturbed operator is affected.

We should also consider the single-trace operators $j^t$ as a potential relevant perturbations.   We consider first the case that this is forbidden by a symmetry and the double-trace is the leading relevant term.  This is somewhat unnatural for the charge density, though it may apply to spin density as we discuss in Sec.~3.3.  In 3.2.2 we look at the single-trace perturbation.

\subsubsection{Density bilinears}

Let focus on a density-squared perturbation,
\begin{equation}
S_{\rm pert} = -\frac{\alpha}{2} \int dt \frac{d^2 k}{(2\pi)^2} j^t_{\vec k} j^t_{-\vec k} \,.
\label{jj}
\end{equation}
Higher powers of the density are also relevant but do not affect the two-point correlator in the planar limit.
The dimension zero case is rather degenerate, so let us consider a general Lifshitz exponent $z$.  As we reduce $z$ the number of relevant interactions decreases, the energy dimension of the current being $2/z$ and the condition for relevance being total dimension less than $1+ 2/z$. Thus the density-squared perturbation is relevant for $z > 2$~\cite{HPST} (the marginal case $z=2$ is left for the reader).  The charge density-density correlator in the strongly coupled theory is of the form
\begin{equation}
{\cal G}^{j}_0(\vec k,\omega)  = k^{2-z} f(\omega^2/k^{2z}) = \omega^{-1+2/z} \tilde f(k^2/\omega^{2/z})  \, .\label{jjex}
\end{equation}
The same summation that gave the fermion propagator~(\ref{gcorr}) gives the full density-density correlator
\begin{equation}
{\cal G}^{j}_{\alpha}(\omega)  = \frac{{\cal G}^{j}_{0}(\vec k,\omega)}{1 + \alpha {\cal G}^{j}_{0}(\vec k,\omega)}\,.
\end{equation}
For $z<2$ this flows to the unperturbed~(\ref{jjex}).  However, for $z>2$ the low energy behavior is
\begin{equation}
{\cal G}^{j}_{\alpha}(\vec k,\omega)  = \frac{1}{\alpha} - \frac{1}{\alpha^2 {\cal G}^{j}_{0}(\vec k,\omega)} + \ldots \,.
\end{equation}
The leading piece is a contact term.  The leading nonlocal behavior corresponds to the dimension of the density being shifted to $1$, and so the density-density perturbation around this new fixed point is irrelevant.

This is suggestive.  At positive $\alpha$ the charge fluctuations in the IR have a certain behavior. When $\alpha$ is reduced to zero, these fluctuations become stronger in the IR.  Once $\alpha$ goes negative, the correlator has a pole at nonzero $k$ for zero $\omega$, indicating an instability.  The enhanced fluctuations at $\alpha=0$ may be connected with the critical behavior.  If so, this is a somewhat conventional picture~\cite{hertzmillis}.  The holography is playing a role in determining the correlators in the strongly coupled theory, where the density fractionalizes~\cite{SVBSF} as in the earlier discussion.  (One might hope that the bulk description will determine the endpoint of the $\alpha < 0$ instability, although the details here are not clear).  However, in the end we must treat by hand the interactions between the singlet fluctuations and the fermions, unless we can find a more effective dual theory.

As a technical point, it might seem puzzling that a density such as $ j^t$ could acquire an anomalous dimension, because the Ward identity appears to fix its dimension via the OPE
\begin{equation}
\omega j^t(\vec k, \omega) \Psi_s - \vec k \cdot \vec\jmath(\vec k,\omega) \Psi_s
= i \Psi_{s}\ .
\end{equation}
One can perturb the terms in this relation as we have done above, with the result
\begin{eqnarray}
j^t(\vec k, \omega) \Psi_s &\to& \frac{1}{1 + \alpha{\cal G}^{j}_{0}(\vec k,\omega)}j^t(\vec k, \omega) \Psi_s|_{\alpha = 0} \, ,\nonumber\\
\vec \jmath(\vec k, \omega) \Psi_s &\to& \vec \jmath(\vec k, \omega) \Psi_s|_{\alpha = 0} 
-  \frac{\alpha{\cal G}^{j}_{0}(\vec k,\omega)}{1 + \alpha{\cal G}^{j}_{0}(\vec k,\omega)}
\frac{\vec k \omega}{k^2} j^t(\vec k, \omega) \Psi_s|_{\alpha = 0}  \,.
\end{eqnarray}
The Ward identity is still satisfied, but for $z > 2$ the $\omega j^t(\vec k, \omega) \Psi_s$ makes no contribution in the IR and so the dimension of $j^t$ is not constrained.  Rather, the $1/k^2$ term, representing an outflow of charge at spatial infinity, appears in the IR and allows the spatial current term by itself to saturate the Ward identity.  One can give this a physical interpretation as follows.  In Ref.~\cite{HPST}, it was noted that the IR stable quantization for $z > 2$ forbids local fluctuations of the charge density (where in the present discussion charge is replaced by spin).  When a nonsinglet operator such as $\Psi_{s}$ acts to create a local density, this must be accompanied by an outflow at infinity.

\subsubsection{Linear density terms}

The current $j^t$ is always relevant according to the standard dimensional analysis, given below Eq.~(\ref{jj}).  This would have a large effect on the IR: the strongly coupled CFT has charged excitations of arbitrarily low energy, so adding a bulk chemical potential $\mu' j^t$ to the action will induce a density of these charges.  In fact, in the $AdS_2$ and Lifshitz geometries 
of \cite{Goldstein:2009cv} this density should already be present: the $AdS_2$ geometry is sourced by a charged horizon, and the Lifshitz geometry in \cite{Goldstein:2009cv} is sourced by a charge density in the bulk.  (We could consider models with multiple $U(1)$'s, but here we are looking only at the simplest case).   

Thus we should expand
\begin{equation}
j^t = \langle 0 | j^t | 0 \rangle + \delta j^t \ .
\end{equation}
The leading contribution to the current is proportional to a $c$-number in the IR.  Assuming that the classical $AdS_2$ or Lifshitz solution is an IR attractor, the leading operator contribution $\delta j^t$ will be irrelevant.  When there is a bulk charge density, the mass of the bulk photon is lifted by Higgsing or plasma effects, and with it the dimension of $\delta j^t$.  In the $AdS_2$ case, with the Einstein-Maxwell action, mixing between the photon and the metric lifts the dimension of $\delta j^t$ to 2.

We present here the derivation of this last result.
We take the background Ansatz 
\begin{equation}
L^{-2} ds^2 =  dy^2  - e^{2\gamma_t(y)} dt^2  + e^{2\gamma_x(y)} dx^i dx^i \, ,
\end{equation}
with vector potential $A_t(y)$.  The Einstein-Maxwell field equations (in units where the $U(1)$ coupling is 1) are
\begin{eqnarray}
0 &=& 1 - 4\ddot \gamma_x - 6 \dot \gamma_x^2 - e^{-2\gamma_t} \dot A_t^2
\nonumber\\
&=& 1 - 2\ddot \gamma_x - 2\ddot \gamma_t - 2 \dot \gamma_x^2 -2 \dot \gamma_x \dot \gamma_t  - 2 \dot \gamma_t^2 + e^{-2\gamma_t} \dot A_t^2
\nonumber\\
&=& 1 - 4 \dot\gamma_t \dot \gamma_x - 2 \dot \gamma_x^2 - e^{-2\gamma_t} \dot A_t^2
\nonumber\\
&=& e^{-\gamma_t} \ddot A_t -  e^{-\gamma_t} (\dot \gamma_t - 2\dot \gamma_x) \dot A_t \,;
\end{eqnarray}
a dot denotes $\partial_y$.
Linearizing around the $AdS_2$ solution,
\begin{equation}
\gamma_t = - y + \delta\gamma_t \,,\quad \gamma_x = \delta\gamma_x \,,\quad e^{-\gamma_t} \dot A_t = 1 + \varphi  \,,
\end{equation}
the field equations become
\begin{eqnarray}
0 &=& - 4\dot u_x - 2 \varphi
\nonumber\\
&=& - 2\dot u_x - 2\dot u_t + 2 u_x  +4 u_t + 2\varphi
\nonumber\\
&=& 4 u_x - 2\varphi
\nonumber\\
&=& \dot\varphi + 2 u_x \,,
\end{eqnarray}
in terms of the derivatives $u_{t,x} ={\delta \dot\gamma}_{t,x}$.  The two solutions
\begin{eqnarray}
u_t &=& e^{2y} \,, \quad u_x = \varphi = 0 \,, \nonumber\\
u_t &=& 4 e^{-y}\,, \quad u_x = 3 e^{-y} \,, \quad \varphi = 6 e^{-y} \,,
\end{eqnarray}
scale respectively as the normalizable and nonnormalizable modes for $\Delta = 2$ in a 1+0 dimensional CFT. This should be compared to the universal result on the 
anomalous scaling dimension of
the vector part of the current at zero momentum: 
$\Delta = 2$, found in \cite{Edalati:2009bi} and studied
further in the transverse channel at non-zero momentum in \cite{Edalati:2010hk}.  The non-zero $k$ analysis for the longitudinal part of the current would be valuable to compare to the above result. 

In summary, there is generically no $U(1)$ charge density of canonical dimension in the IR CFT, and the Ward identity must be satisfied without it as in the discussion in Sec.~3.2.1.  Perturbations involving the $U(1)$ density in the UV do not destabilize the IR physics, with one notable exception: the relativistic case $z=1$ is unnatural for a CFT with a $U(1)$ charge, because the relativistic symmetries are broken by any charge density.

\subsection{Spin-orbit coupling}

There is another set of relevant operators that we can add to
(\ref{action}) which we have thus far neglected. 
These are the spin-orbit Rashba \cite{rashba} terms,\begin{equation}
S_{SO} = \int dt \frac{d^2 k}{(2\pi)^2} \left(
r_{\vec k} \chi^\dagger_{\vec k } ( \epsilon_{ij} \sigma^i k^j ) \chi_{\vec k}
+ u_{\vec k} \chi^\dagger_{\vec k } ( \epsilon_{ij} \sigma^i k^j ) \Psi_{\vec k} +
u_{\vec k}^* \Psi^\dagger_{\vec k } ( \epsilon_{ij} \sigma^i k^j ) \chi_{\vec k}
\right)
\label{so}
\end{equation}
where $\sigma^i, \, i = 1,2$ are the usual Pauli matrices in the spin basis $s$.  These terms respect two-dimensional parity, as compared to a term such as
as $\chi^\dagger \sigma^i k^i \chi$ which does not.\footnote{Recall that parity in two space dimensions
is defined by reflection about a single axis, and as usual $\chi^\dagger \sigma^i \chi$ is a pseudovector under this parity.}    For a two-dimensional system embedded in three dimensions, the Rashba interaction arises from an interaction~$\epsilon_{ijk} \sigma^i k^j \hat z^k$.  Treating $\hat z$ as a vector, this respects the underlying parity symmetry but requires a spontaneous breaking of reflection symmetry in the third direction, for example by the crystal structure.   When this symmetry is broken these terms can have interesting physical consequences \cite{spinhall}.

Allowing the terms (\ref{so}) will be important for us because they are present 
in the $AdS_4$ theory, and must be included
in order to match to that theory.  In that case the reflection symmetry in the third direction is broken by the warped geometry.  In particular, as we
will explain below, because the strongly coupled field $\Psi$
naturally does not respect this ``reflection'' symmetry it would constitute a fine tuning
not to include these terms.

We can diagonalize the full action in terms of the eigenbasis under reflection perpendicular to the momentum,
\begin{equation}
|\alpha\rangle = e^{-i\theta/2} |\uparrow\rangle + (-1)^{\alpha+1}  e^{i\theta/2} |\downarrow \rangle\,,\quad \alpha=1,2 \,,
\end{equation}
where $e^{i\theta}$ is the phase of $k^1 + ik^2$.  
This is the basis that Ref.~\cite{Faulkner:2009wj} works in. Then, 
\begin{equation}
S + S_{SO} = S_{\rm strong} +  \sum_\alpha \int dt\int \frac{d^2 k}{(2\pi)^2} \, \left\{ \chi_{\vec k,\alpha}^\dagger (i \partial_t -  \varepsilon_{\vec k}^\alpha + \mu) \chi^{\vphantom\dagger}_{\vec k,\alpha}
+ g^{\alpha \vphantom *}_{\vec k} \chi_{\vec k,\alpha}^\dagger \Psi^{\vphantom\dagger}_{\vec k,\alpha} + g_{\vec k}^{\alpha *}  \Psi^{\dagger}_{\vec k,\alpha} \chi_{\vec k,\alpha}^{\vphantom\dagger} \right\} 
\end{equation}
where,
\begin{equation}
\label{eq:rashba}
\varepsilon_{\vec k}^\alpha =  \varepsilon_{\vec k} + (-1)^{\alpha+1}
|k| r_{\vec k},  \qquad
g^\alpha_{\vec k} = g_{\vec k} +  (-1)^{\alpha+1} |k| u_{\vec k} \,.
\end{equation}
This term splits the degenerate Fermi surfaces labeled by spin $s$
into two non degenerate fermi surfaces labeled by $\alpha$. For example
in this basis the generalization of (\ref{gcorr}) is found by replacing 
$v_{\vec k_\star} k_\parallel \rightarrow v_{\vec k_\star}^\alpha k_\parallel^\alpha$
with $k_\parallel^\alpha = | \vec{k} - \vec{k}_\star^\alpha|$.  
Note that time reversal is still preserved, and simply maps $\vec{k}, \alpha$ to $-\vec{k},\alpha$.

In addition, in this basis the correlator coming from the strongly coupled sector for the $\Psi$ field will be diagonal since we demand that the strongly coupled
 field theory respects $T$ and $P$. That is
in (\ref{gcorr}) we should replace $c_k \rightarrow c_k^\alpha$. There
is no reason for this sector to respect the ``reflection'' symmetry discussed above, thus
$c_k^1 \neq c_k^2$ and in the original spin basis the correlator will not be diagonal.
Note that in the supergravity approximation the dimension of $\Psi$ is independent of $\alpha$, but no symmetry appears to require this at higher order.

Aside from the reflection symmetry, spin-orbit interactions are also forbidden in the extreme nonrelativistic limit, where $SU(2)_{\rm spin}$ emerges as an accidental symmetry (as in heavy quark systems).  This is often treated as an effective internal symmetry in condensed matter systems.  Thus we might model it by introducing an explicit $SU(2)$ index on the fermionic fields; this could be the original $s$ in Eq.~(\ref{action}), with the additional $\alpha$ index inherited from the bulk field suppressed.  There will be a small symmetry breaking, from the combination of lattice and spin-orbit effects.  

One would then expect that the approximately conserved $SU(2)$ current would be holographically realized as an $SU(2)$ gauge symmetry in the bulk, weakly broken.  The discussion of density bilinears above would extend to the spin density, where the absence of a single-trace term would usually follow from dicrete symmetries.  
  
We can attempt to include some more physics, in particular the fact that the spin fluctuations may be strongest at the antiferromagnetic point.  We thus introduce a lattice as before, elaborating the perturbation to
\begin{equation}
S_{\rm AF} = -\frac{1}{2} \int dt \int_{\rm unit\atop cell}\frac{d^2 k}{(2\pi)^2} \alpha_{\rm s}(k) \sum_{\vec K} j^t_{i,\vec k} j^t_{i,\vec K -\vec k} \,.
\end{equation}
The resulting correlator is 
\begin{equation}
{\cal G}^{\rm s}_{\alpha}(\vec k,\omega)  = \frac{\tilde {\cal G}^{\rm s}_{0}(\vec k,\omega)}{1 + \alpha_{\rm s}(k) \tilde{\cal G}^{\rm s}_{0}(\vec k,\omega)}\,,
\end{equation}
where 
\begin{equation}
\tilde {\cal G}^{\rm s}_{0}(\vec k,\omega) = \sum_{\vec K} \tilde{\cal G}^{\rm s}_0(\vec k + \vec K,\omega)\,.
\end{equation}
In an antiferromagnetic system the coupling $\alpha_{\rm s}(\vec k)$ would be negative at some nonzero momentum.  However (as with the earlier discussion of behavior at the Fermi surface), for any finite $z$ the interesting critical behavior in the bulk is at zero momentum and does not seem to affect the antiferromagnetic point.  
Perhaps we have not introduced the lattice physics in the most general way.  One further possibility would be the introduction of constant $SU(2)$ gauge fields in the bulk.

\sect{Matching onto the UV theory}

In general, the system we have been discussing will describe the low energy physics of some higher energy QFT in 2+1 dimensions.  The UV theory has some leading fermionic operator  $\mathcal{O}_\alpha$ which is represented in the CFT as
\begin{equation}
\mathcal{O}_\alpha(\vec x,t) = \int \frac{d^2 k}{(2\pi)^2} e^{ i \vec{k}.\vec{x}} \left(
Z_{\vk}^\alpha \chi_{\vec{k}, \alpha} (t) + {\mathcal Z}_{\vk}^\alpha \Psi_{\vec{k}, \alpha}(t) + \ldots \right) \,,
\label{uvir}
\end{equation}
where the omitted terms include time derivatives, multitraces, and so on.  Since $\chi_{\vec{k}, \alpha}$ has dimension 0 and $\Psi_{\vec{k}, \alpha}$ dimension at least $\frac{1}{2}$, the leading low energy behavior of the correlator comes from the former, and so from the IR correlator~(\ref{gcorr}) generalized to the $\alpha$ basis we have
\begin{equation}
G^{\mathcal{O}}_{\alpha} (\vec k, \omega) \stackrel{\rm IR}{=} \frac{|Z_{\vec k}^\alpha|^2}{\omega - v_{\vec k_*}^\alpha  k_\parallel^\alpha - c_{\vec k}^\alpha |g_{\vec k}^\alpha|^{2} \omega^{2\Delta_{\vec k}-1} } \,.
\label{uvcorr}
\end{equation}

In the remainder of this section we focus on the case that the UV theory is the original $AdS_4$ theory studied in Refs.~\cite{Lee:2008xf,Liu:2009dm,Cubrovic:2009ye,Faulkner:2009wj}.
We will suppress the index $\alpha$ wherever possible to avoid clutter. 

\subsection{Matching onto the $AdS_4$ theory}

The bulk geometry is the $AdS_4$ (extremal) RN black hole. As a function
of the radial coordinate this geometry represents the RG flow of the field
theory triggered by turning on the relevant charge density operator $\mu j^t_{AdS_4}$
in the UV.  At scales much smaller than $\mu$ the theory flows to the $AdS_2 \times R^2$ theory,
where now $\mu$ represents a UV cutoff.  It seems reasonable to assume that the low energy theory $\omega \ll \mu$ is universally
controlled by the $AdS_2$ theory plus the domain wall singlet fermions.  Here we explore this interpretation.

The fermionic operator $\mathcal{O}(x)$ in the UV CFT can be
considered with general scaling dimension $\Delta_4$ and charge $q$. 
The fermionic correlators can be written as \cite{Faulkner:2009wj},
\begin{equation}
\label{ads4}
G^{\mathcal{O}} (\vec k, \omega) = 
 \mu^{2 \Delta_4-3} \frac{ b_+ + b_- c_{\vec k} (\omega/\mu)^{2\Delta_{\vec k} -1 }}{ a_+ + a_- c_{\vec k} (\omega/\mu)^{2\Delta_{\vec k} -1 }} 
\end{equation}
where
\begin{equation}
a_\pm(k,\omega,\mu) = \sum_n  (\omega/\mu)^n a_{\pm}^{(n)}(k/\mu) \qquad 
b_\pm(k,\omega,\mu) = \sum_n (\omega/\mu)^n  b_{\pm}^{(n)}(k/\mu) \,.
\end{equation}
and $a_\pm^{(n)}, b_\pm^{(n)}$ are only calculable
numerically order by order in perturbation theory.
The IR dimension is fixed by the specific $AdS_2 \times R^2$ theory to be \cite{Faulkner:2009wj},
\begin{equation}
\label{exponent}
\Delta_{\vec{k}} = \frac{1}{2} + \sqrt{ \frac{(\Delta_4-3/2)^2}{6} + g_F^2\left( \frac{k^2}{2 \mu^2} - \frac{q^2}{12} \right) } \,,
\end{equation}
where $g_F$ is an order one number fixed by the normalization of the
current two point function. It would be useful to understand how
robust this relation is, or whether there are other $AdS_2$ theories
which give different scaling dimensions.

As discussed in \cite{Faulkner:2009wj} the condition for the appearance of a
Fermi surface is $a_+^{(0)}(k_\star/\mu) = 0$.  Expanding at low energy, this matches the general result~(\ref{uvcorr}) up to a contact term, with the identifications
\begin{eqnarray}
|Z_k|^2 &=&
\mu^{2\Delta_4-2} b_+^{(0)}/ a_+^{(1)} \,,\nonumber\\
\varepsilon_{\vec k} - \mu &=& \mu a_+^{(0)}/a_+^{(1)} \,,\nonumber\\
|g_{\vec k}|^2 &=& - \mu^{-2\Delta_{\vec k}+2} a_-^{(0)}/a_+^{(1)} \,.
\end{eqnarray}
To reproduce higher order terms in (\ref{ads4}) we would have to include
general time derivatives in the expansion~(\ref{uvir}), and also go to higher order in the expansion~(\ref{linear}) of the low energy action.

For $\Delta_{\vec k} < 1$ the observation at the end of section 3.1 gives a nice
picture of the RG flow represented by the $AdS_4$ theory. Since we can
now ignore the free field $\chi_{\vec k}$ in favor of the interacting field 
$\Psi_{\vec k}$ in the IR CFT, we can match directly onto the $AdS_2 \times R^2$
theory via the strongly interacting fermionic operator in that theory; 
\begin{equation}
\mathcal{O}(x) = \int \frac{d^2 k}{(2\pi)^2} e^{ i \vec{k}.\vec{x}}
\tilde{\cal Z}_{k} \Psi_{\vec{k}} (t) + \ldots \,, \quad |\tilde{\cal Z}_k|^2 = 
- \mu^{2(\Delta_4-\Delta_{\vec k}-1)}  \frac{b_+^{(0)} a_-^{(0)}}{ (a_+^{(0)})^2}  \,.
\end{equation}
We must also specify the presence of an irrelevant double-trace deformation as in (\ref{fm}),
\begin{equation}
\label{multi}
\frac{1}{f_{\vec k}} = \frac{ |g_{\vec k}|^2}{\mu - \varepsilon_{\vec k} }
= - \mu^{1-2\Delta_{\vec k}} \frac{ a_-^{(0)}}{ a_+^{(0)}}\, .
\end{equation}
after which the $\Psi$ correlator 
is given by the earlier result~(\ref{intout2}). 

It is natural that  under RG flow all multi-trace relevant operators in the IR theory not protected
by a symmetry should be turned on. Hence we expect naturally the IR theory is
in its standard quantization.
However, we have seen in Sec.~3.1 that it is precisely on the Fermi surface that we are tuned to the alternate quantization, and the IR fluctuations are enhanced.  Generically this tuning will happen on a codimension one surface in momentum space.  As we move away from the Fermi surface, the alternate quantization holds only down to a crossover scale $\Lambda_{\vec k} = \mu ( a_+^{(0)}/a_-^{(0)})^{1/(1-2\Delta_{\vec k})} \sim \mu (k_\parallel/\mu)^{ 1/(1-2\Delta_{\vec k} )}$.  

\begin{figure}[h!]
\begin{center}
\includegraphics[scale=.6]{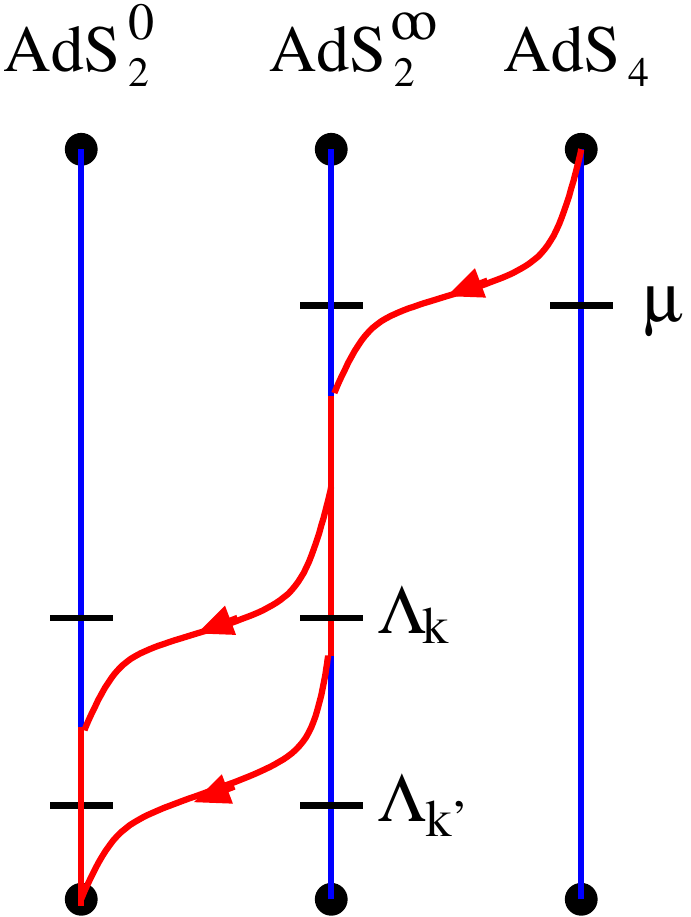} \hspace{1cm}
\includegraphics[scale=.9]{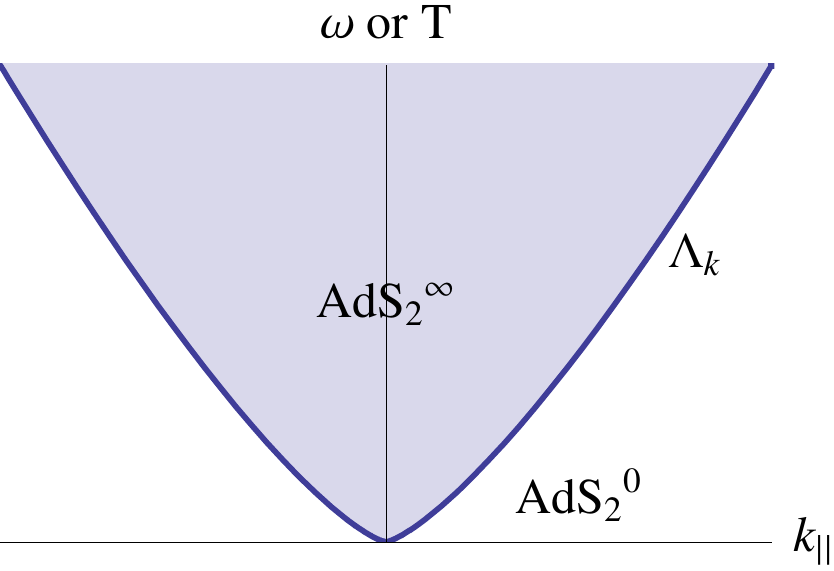}
\caption{ 
Left: flows represented by the $AdS_4$ RN geometry in the presence of
a probe fermion when $\Delta_{\vec k} < 1$. The vertical direction is an energy scale and $\Lambda_k > \Lambda_k'$ for
the two different flows. The standard (alternative) quantized theory is
denoted $AdS_2^0$ ($AdS_2^\infty$).
Right: the fermi surface with $\Delta_{\vec k} < 1$
is a quantum critical point in $k$ space, with $AdS_2^\infty$ controlling
the physics.
\label{fig:1}  }
\end{center}
\end{figure}

This interpretation is depicted in Fig.~\ref{fig:1}.  We flow in the UV theory (denoted $AdS_4$) to the theory at the UV end of the IR CFT (denoted $AdS_2$).  This IR CFT has an infinite number of double-trace couplings, indexed by $\vec k$.  For  $\vec k$ near the Fermi surface, the coupling lingers for a long time near the unstable fixed point corresponding to the alternate quantization, while values of $\vec k$ further from the Fermi surface cross over sooner.
The $AdS_2$ CFT
with alternative quantization is controlling the physics of the Fermi surface. This 
interpretation explains the universal behavior recently found for the transport  \cite{FILMV} 
and thermodynamics \cite{Hartnoll:2009kk,Denef:2009yy}
of this model when $\Delta_{\vec k} < 1$. 

Such an interpretation for $\Delta_{\vec k} > 1$ is hampered by the now relevance of
the time derivative term in (\ref{action}). Thus the low energy effective theory
is no longer just the $AdS_2$ theory. Indeed, close to the fermi surface the
theory is now controlled by the free fermion $\chi$. 

It would be interesting to match the low energy gauge field sector of the $AdS_4$ theory to the discussion of bosonic bilinears in the last section. This may help
to track the instability identified for the density bilinear. We leave this to future work.

\subsection{Relevant and irrelevant deformation of the $AdS_4$ theory}

The space of possible low energy theories (\ref{action}) that
can be constructed with this $AdS_4$ theory was examined
in Ref.~\cite{Faulkner:2009wj}. Here we generalize these results by allowing for double-trace
fermion operators in the $AdS_4$ theory. We start by considering
relevant operators, noting that there is potentially another natural candidate
for the doping. However we then note that since the UV CFT is not
governing the low energy physics the distinction between irrelevant
and relevant in the UV is immaterial. Both should alter the low energy
physics only through their effect on the already identified relevant operators
in the low energy effective action (\ref{action}). 
In particular the cutoff scale $\mu$ above which the $AdS_4$ theory
is important should be at the same scale as the lattice scale and
as such we should include possible lattice effects. We only briefly explore
this possibility.

We will restrict our attention to the region where the bulk fermi sea does
not back react on the $AdS_2$ geometry as in \cite{HPST}. This
condition is equivalent to requiring $\Delta_{\vec k}$ is real for all $k$, or
\begin{equation}
g_F q < |2 \Delta_4 - 3 | \, .
\end{equation}
This corresponds to the regions of Fig.~6 of Ref.~\cite{Faulkner:2009wj}
below the dashed lines. There are only
zeros (shaded areas in that figure) of 
$\epsilon_{\vec k} - \mu \propto a_+^{(0)}$ in the bottom left wedge  of the
$\alpha=2$ component (and none in the $\alpha=1$ component.)
This is when $1/2 < \Delta_{\vec k} < 1$ and 
$1 < \Delta_4 < 3/2$; which is exactly the critical region of the $AdS_2$ theory
and in the alternative quantization of the $AdS_4$ theory. 
%This is somewhat
%of a limited set of theories, for example $\Delta_{\vec k} > 1$ is not realized. 
This means from the UV theory perspective there
are relevant double-trace operators
which when turned on will induce a flow to a theory 
without a Fermi surface \cite{Faulkner:2009wj}.
Here
we extend this observation and note that this flow is quite interesting, in particular
how the Fermi surface disappears depends on the sign of the double-trace coupling.
More specifically we will add to the action the following,
\begin{equation}
S_h = - \int d^3x\, h  \bar{\mathcal{O}} \gamma^t \mathcal{O}^{\vphantom \dagger} (x) =  \sum_\alpha \int dt \frac{d^2k}{(2\pi)^2}  h\,  \mathcal{O}_\alpha^\dagger(\vec{k},t) \mathcal{O}_\alpha^{\vphantom \dagger}(\vec{k},t)\,.
 \end{equation}
which is like a chemical potential for $\mathcal{O}$ distinct from $\mu$.
Note one could also add the relevent, Lorentz invariant term $\bar{\mathcal{O}} \mathcal{O}$, but this violates $P$ and $T$ in $2+1$ dimensions, so we will not include it.
We will consider the dimensionless quantity $\mathtt{x}  \equiv h \mu^{-3+2 \Delta_4}$. 

Suppressing
$\alpha$ dependence, we can sum the double-trace perturbations as before, 
\begin{equation}
G^{\mathcal{O}}_{h}(\vec k, \omega) 
 = \frac{ G^{\mathcal{O}}(\vec k, \omega) }
 {1 + h G^{\mathcal{O}}(\vec k, \omega) }
  = \mu^{2 \Delta_4-3} \frac{ b_+ + b_- c_{\vec k} (\omega/\mu)^{2\Delta_{\vec k} -1 }}{ ( a_+
  + \mathtt{x} b_+)  + (a_- + \mathtt{x} b_- ) c_{\vec k} (\omega/\mu)^{2\Delta_{\vec k} -1 }} \,.
\end{equation}
There is now a zero-frequency pole of the propagator at
\begin{equation}
0 = a_+^{(0)}(k_\star/\mu) + \mathtt{x} b_+^{(0)}(k_\star/\mu) \propto \varepsilon_{\vec k} - \mu 
\end{equation}
As 
$\mathtt{x}$ increases the fermi momentum $k_\star$ increases and diverges as $\mathtt{x} \rightarrow \infty$.  Decreasing $\mathtt{x}$ the fermi surface vanishes
beyond some point $\mathtt{x} < \mathtt{x}^c $.
Indeed $\mathtt{x}$ may be a candidate
for the doping although no new critical behavior will happen
as a function of $\mathtt{x}$ and there is nothing to single out $\Delta_{\vec k} = 1$
as special.

\begin{figure}[h!]
\begin{center}
\includegraphics[scale=1.2]{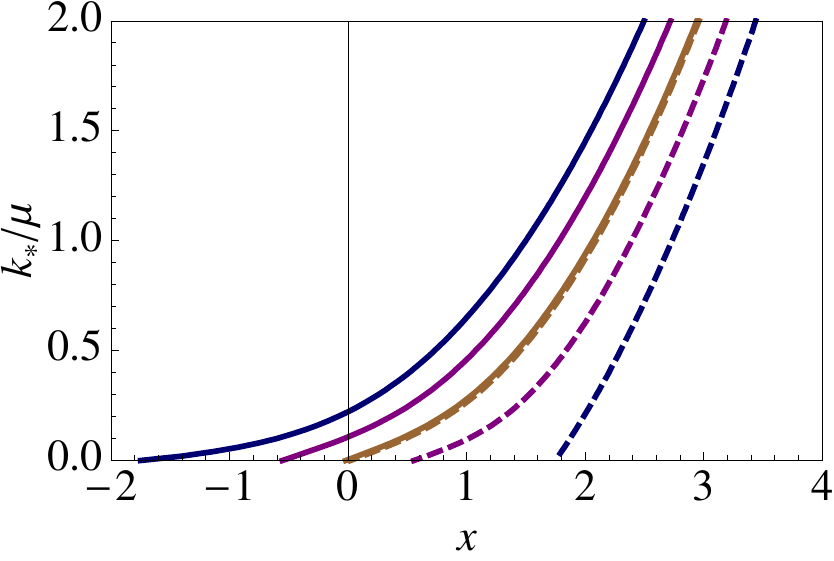}
\caption{ 
 \label{fig:2} Size of the fermi surfaces $k_\star^{\alpha=2}$ (solid) and $k_\star^{\alpha=1}$ (dashed) as a function
of the relevant double-trace coupling $\texttt{x}$ in units of $\mu$ for
three choices of $q$ and $\Delta_4 = 5/4$. The splitting due
to spin orbit coupling increases with $q$, and is absent here for $q=0$.
Note that the IR scaling dimension $\Delta_{\vec k_\star}$ increases with ``doping''
because it increases with $k_\star$.}
\end{center}
\end{figure}

In Figure~\ref{fig:2} some flows under $\mathtt{x}$ were constructed explicitly with
the $AdS_4$ theory. Note that the $AdS_2$ exponent $\Delta_{\vec k_\star}$ 
increases as a function of $\mathtt{x}$; for fixed $q,\Delta_4$ it now varies
from $1/2 + \sqrt{ (\Delta_4-3/2)^2 - g_F^2 q^2/2 }/\sqrt{6} < \Delta_{\vec k} < \infty$, a much
larger parameter space than discussed earlier. 
It is interesting to note from the middle curve in Figure~\ref{fig:2}, which has $q=0$,
that there can now exist fermi surfaces 
(of non-zero size) when the fermion is uncharged. Also for $q=0$ the
spin orbit terms (\ref{so}) are naturally zero, presumably due to a particle-hole symmetry.
One thing
to take away from this section is that the size of the Fermi surface in the $AdS_4$ 
model can be tuned independently of $\mu$. This will be useful in
isolating and studying the thermodynamic properties of the Fermi surface,
including their response to an external magnetic field already studied in \cite{Denef:2009yy,magnetic}.\footnote{We
thank Hong Liu for emphasizing this point.}

Notice there is nothing special about the point $\texttt{x}=0$. The reason for
this is the UV CFT is not controlling the physics of the fermi surface. Indeed
since $k_\star$ is at the same scale as the cutoff scale $\mu$,  the lattice size 
should also be at the scale $a \sim 1/\mu$.   
So there is no reason not to include
general irrelevant and relevant deformations at this scale, ones
that break translation and rotation invariance, to mimic the effects
of a lattice. This is easier to achieve using double-trace deformations, although
there is no reason to not turn on single-trace operators; it is just much harder to analyze.

Consider adding to the $AdS_4$ theory the $h$ deformation but
now 
modulated on the scale of the lattice, $h(x)$.
Then the two point function can be shown to be,
\begin{equation}
G^{\mathcal{O}}_{h(x)}(\vec k, \omega) 
 = \frac{ G^{\mathcal{O}}(\vec k, \omega) }
 {1 + \sum_{\vec K} h_{\vec K} G^{\mathcal{O}}(\vec k+ \vec K, \omega) }
\end{equation}
As above we can match this theory onto (\ref{action}) where we must
now restrict the momentum $\vec{k}$ 
to the unit cell of the reciprocal lattice, and allow for a coupling as in (\ref{umk}).
For example the condition for a fermi surface is,
\begin{equation}
\varepsilon_{\vec k} - \mu \propto a_+^{(0)} (k/\mu) \left( 1 + \sum_{\vec K} h_{\vec K} \mu^{2 \Delta_4 -3} 
 \frac{ b_+^{(0)}(|\vec{k} + \vec{K}|/\mu)}{a_+^{(0)}(|\vec{k} + \vec{K}|/\mu)} \right) = 0 \,.
\end{equation}
Hence this construction sits within the universality class of low energy theories
in (\ref{action}). Although this construction on its own might be useful for analyzing the
transport and thermodynamic properties of (\ref{action}).

%======================================================================================

\section*{Acknowledgments}
We would like to thank S. Hartnoll, N. Iqbal, H. Liu, J. McGreevy, E. Silverstein, D. Tong, and D. Vegh for extensive discussions and collaboration on related questions.  We also thank L. Balents, S. Kachru, H. Katsura, V. Kumar and M. Roberts for useful discussions and E. Silverstein for suggestions on the manuscript.  JP is grateful to the Institute for Advanced Studies for hospitality while part of this work was carried out.  This work was supported in part by NSF grants PHY05-51164 and PHY07-57035 and the UCSB physics department. 

%================================BIBLIOGRAPHY=====================================

\end{document}